\documentstyle[aps,twocolumn,prl,epsf]{revtex}

\begin{document}

\newcommand{\be}{\begin{equation}}
\newcommand{\ee}{\end{equation}}

\draft

\twocolumn[\hsize\textwidth\columnwidth\hsize\csname @twocolumnfalse\endcsname

\title{Electron Energy Loss Spectroscopy of strongly correlated systems 
in infinite dimensions}
\author{Luis Craco$^1$ and Mukul S.Laad$^2$}

\address{${}^1$Instituto de Fisica ``Gleb Wataghin'' - UNICAMP, C.P. 6165, 
13083-970 Campinas - SP, Brazil \\ 
${}^2$Institut fuer Theoretische Physik, Universitaet zu Koeln, Zuelpicher 
Strasse, 50937 Koeln, Germany  
}
\date{\today}
\maketitle

\widetext

\begin{abstract}
We study the electron-energy loss spectra of strongly correlated electronic
systems doped away from half-filling using dynamical mean-field theory 
($d=\infty$).  The formalism can be used to study the loss spectra in the 
optical (${\bf q=0}$) limit, where it is simply related to the optical 
response, and hence can be computed in an approximation-free way in 
$d=\infty$. We apply the general formalism to the one-band Hubbard model 
off $n=1$, with inclusion of site-diagonal randomness to simulate effects of 
doping. The interplay between the coherence induced plasmon feature and 
the incoherence-induced high energy continuum is explained in terms of the 
evolution in the local spectral density upon hole doping. Inclusion of static 
disorder is shown to result in qualitative changes in the low-energy features, 
in particular, to the overdamping of the plasmon feature, resulting in a 
completely incoherent response. The calculated EELS lineshapes are compared 
to experimentally observed EELS spectra for the normal state of the 
high-$T_{c}$ materials near optimal doping and good qualitative agreement 
is found.
\end{abstract}
\pacs{PACS numbers: 75.30.Mb, 74.80.-g, 71.55.Jv}

]

\narrowtext

The ac conductivity and dielectric tensor provides valuable information 
concerning the finite frequency, finite temperature charge dynamics of an
electronic fluid in a metal. The dramatic changes in the electronic state 
from localized to itinerant across an insulator-metal (I-M) transition, be 
it driven by pressure, or doping the insulator, is reflected in concomitant 
changes in the optical responses. These studies, therefore, give one a 
systematic picture describing the nature of the change in electronic state, 
appearance of new low-energy excitations, their dispersion and stability.  
In cuprate superconductors, for e.g, these studies have convincingly 
demonstrated the non-Fermi liquid nature of the charge~\cite{[1]} and 
spin~\cite{[2]} dynamics in the normal state. 

  Optical response in a solid provides an estimate of the carrier scattering 
rate at finite frequency, and thus give detailed information about the finite
frequency collective excitations responsible for scattering the carriers.  In
a quantum paramagnetic metal, the dominant carrier-scattering mechanism are
electron-hole pairs, which are the low-energy collective excitations- 
it is precisely the e-h spectrum which   
is measured by electron-energy-loss spectroscopy (EELS), which provides 
one with a direct experimental picture for the spectral density of 
particle-hole excitations in a solid~\cite{[3]}.

  In a weakly interacting system, one expects the details of the particle-hole
spectrum to be sensitive to details of the shape and size of the Fermi surface
(e.g, its curvature).  That such a picture is untenable for strongly 
correlated
systems was pointed out by Shastry {\it et al}~\cite{[4]}, who argued that in 
this case, the whole Brillouin zone tends to get populated.  Another 
important point of considerable relevance is the transfer of spectral weight 
over large energy scales that is characteristic of strongly correlated 
systems-it has no analog in weakly interacting fermi systems.  Consequently, 
one expects qualitatively different responses in this case, compared to those 
characteristic of weakly interacting systems.  

 Quite generally, the transmition EELS lineshape is related to the 
wave-vector and frequency dependent dielectric function via the equation,
\be
\label{eq:iqw}
I({\bf q},\omega)=-Im\frac{1}{\epsilon({\bf q},\omega)} \;.
\ee
Following Ref.~\cite{[3]}, we employ the random-phase-approximation (RPA) 
to treat the small-${\bf q}$ part.  In the optical limit that we are 
interested in ($q\rightarrow 0$), 
$\epsilon({\bf q}\rightarrow 0,\omega)=1+4\pi\frac{i\sigma(\omega)}{\omega}$, 
where $\sigma(\omega)$ is the longitudinal optical conductivity.
The analysis carried out along these lines is invalid when $qa \simeq 1$, 
where the short-ranged part of the potential has to be considered and the 
RPA is inadequate.  However, local correlation effects are correctly 
incorporated into the above eqn via $\sigma(\omega)$, which can be computed 
reliably within $d=\infty$, for example. In this limit, ${\bf q} \simeq 0$, 
the particle-hole spectral density is obtainable from the dissipative part 
of~\cite{[3]},
\be
I(0,\omega)=-Im\frac{1}{\epsilon(0,\omega)}=
\frac{\epsilon"(\omega)}{\epsilon'^{2}(\omega)+\epsilon"^{2}(\omega)}
\ee
with $\epsilon'(\omega)=1-\frac{4\pi\sigma"(\omega)}{\omega}$ and 
$\epsilon"(\omega)=\frac{4\pi\sigma'(\omega)}{\omega}$
so that one needs to have an estimate of the longitudinal dielectric constant 
to calculate the EELS spectrum.  It is to be noted that this yields the EELS 
lineshape corresponding to the experiment performed in the so-called
"transmission mode".  Thus, the problem has now been reduced to that
of computing $\epsilon(\omega)$, which, being a local quantity, can be computed
in an approximation-free way in $d=\infty$~\cite{[5]}.

In this communication, we want to develop a theory of the trasmition 
EELS (eq.~(\ref{eq:iqw})) for electronic 
systems which undergo Mott transitions as a function of 
band-filling~\cite{[5]}. To be specific, we consider the one-band Hubbard 
model, where the Mott transition can be driven by hole-doping the Mott 
insulator, or by pressure (which changes the ratio $U/t$). To capture more 
fully the effects of chemical substitution-(which is how doping is carried 
out in practice), we also introduce static, site-diagonal disorder in the 
model.  The resulting hamiltonian is,
\be
\label{model}
H=-\sum_{ij\sigma}t_{ij}c_{i\sigma}^{\dag}c_{j\sigma} + U\sum_{i}
n_{i\uparrow}n_{i\downarrow} - \sum_{i,\sigma}(v_{i}-\mu)n_{i\sigma}
\ee
as a prototype model describing the electronic degrees of freedom in TM 
oxides. To study the 3D case, we employ the $d=\infty$ approximation, 
which is the best among those currently available to study the M-I 
transition~\cite{[5]}.  Since this method has been extensively reviewed, 
we only summarize the relevant aspects. All transport properties, which 
follow from the conductivity tensor, are obtained from a ${\bf k}$-independent 
self-energy in $d=\infty$; the only information about the lattice structure 
comes from the free band dispersion in the full Green fn:
\be
\label{gk}
G(k,\omega)=G(\epsilon_{k},\omega)=\frac{1}{\omega+\mu-\epsilon_{k}
-\Sigma(\omega)}
\ee
To solve the model in $d=\infty$ requires a reliable way to solve the single 
impurity Anderson model(SIAM) embedded in a dynamical bath described by the
hybridization fn. $\Delta(\omega)$.  There is an additional condition that 
completes the selfconsistency:
\be
\int d\epsilon G(\epsilon, i\omega)\rho_{0}(\epsilon) = 
\frac{1}{i\omega+\mu-\Delta(i\omega)-\Sigma(i\omega)}
\ee
where $\rho_{0}(\epsilon)$ is the free DOS ($U=0$).  The above 
eqns.~(\ref{model}-\ref{gk}) 
refer to the disorder-free Hubbard model.  With microscopic, binary disorder
(which is the only type we consider here), 
\be
P(v_{i})=(1-x)\delta(v_{i})+x\delta(v_{i}-v)
\ee
only the disorder-averaged quantities are physically observable.  In this case,
we use an extended version of the iterated perturbation theory (IPT), which 
combines the effects of dynamical correlations (via usual IPT) with those of 
disorder (via CPA) in a selfconsistent way.
For the sake of completeness, we outline the calculational details briefly.
As a first step, we compute the full local Green function for the pure model
using IPT.  This has been shown~\cite{[5]}
  to yield results in excellent agreement
with those obtained from exact diagonalization techniques.  This IPT propagator
is then corrected for repeated scattering from local disorder by computing the
new self energy from the usual CPA ~\cite{[6]},
\be
<T_{ii}[\Sigma(\omega)]>_{c}=\frac{-(1-x)\Sigma(\omega)}{1+\Sigma(\omega)G(\omega)} + 
\frac{x(v-\Sigma(\omega))}{1-(v-\Sigma(\omega))G(\omega)}=0
\ee
which results in an implicit equation for the interaction and disorder corrected
 self energy (and Green function).  To make the treatment selfconsistent, we
compute the new IPT selfenergy using the new GF computed with (IPT+CPA) in the
first step.  This procedure is repeated until convergence is achieved.  At each
step of the iteration, we fix the chemical potential from the Luttinger sum 
rule, $n=\int_{-\infty}^{\mu}\rho(\omega)d\omega=x$.
 This extended IPT yields the 
local propagator, $<G_{ii}(\omega)>$ corrected both for dynamical correlations 
and disorder induced repeated scattering, both treated on the same 
footing~\cite{[6]}. In $d=\infty$, this is sufficient to compute the 
transport, because the vertex corrections in the Bethe Salpeter eqn. for 
the conductivity vanish identically in this limit~\cite{[7]}.  Thus, the 
conductivity is fully determined by the basic bubble diagram made up of 
fully interating local GFs of the lattice model.

The optical conductivity and the Hall conductivity are computable in terms 
of the full $d=\infty$ GFs as follows~\cite{[7]}:
\be
\sigma_{xx}(i\omega)=\frac{1}{i\omega}\int 
d\epsilon\rho_{0}(\epsilon)\sum_{i\nu}
G(\epsilon,i\nu)G(\epsilon,i\nu+i\omega)
\ee
The dynamical dielectric constant is directly calculated from the optical 
conductivity via,
\be
\epsilon_{xx}(\omega)=1+\frac{4\pi}{\omega}i\sigma_{xx}(\omega)
\ee
yielding its real and imaginary parts.  Use of eqn.(3) then permits us to 
study 
$I_{EELS}(\omega) = -Im\frac{1}{\epsilon(\omega)}$,

Before embarking on our results and their analysis, it is instructive to 
summarize what is known about the $d=\infty$ Hubbard model.  At large $U/t$, 
and away from half-filling ($n=1$), the ground state is a paramagnetic FL if 
one ignores the possibility of symmetry breaking towards antiferromagnetism,
as well as disorder effects, which are especially important near $n=1$.  This
can be achieved formally by introducing a nnn hopping, which in $d=\infty$
leaves the free DOS essentially unchanged~\cite{[5]}.  This metallic state 
is characterized by two energy scales: a low energy coherence scale $T_{coh}$, 
below which local FL behavior sets in~\cite{[5]}, and a scale of $O(D)$, 
($D$ is the free bandwidth) characterizing high energy, incoherent processes 
across the remnant of the Mott-Hubbard insulator at $n=1$.  At $T<T_{coh}$, 
the quenching of the local moments leads to a response characteristic of a 
FL at small $\omega<<2D$ (but with the dynamical spectral weight transfer 
with doping, a feature of correlations), but at higher $T>T_{coh}$, the 
picture is that of carriers scattered off by effectively local moments, 
which makes the system essentially like a non-FL (note that the
metal with disordered local moments is not a FL).

The picture is modified drastically in presence of static, diagonal disorder,
in which case, the extended IPT has to be used.  For weak disorder, the above
picture is qualitatively unaffected, and the only interesting features are
that Luttinger's theorem is violated, and the FL quasiparticles acquire a 
finite lifetime at $\mu$~\cite{[6]}.  However, with increasing disorder, 
repeated scattering effects destroy the quasiparticle picture, and the 
metallic state is characterized by an incoherent, non-FL response with a 
pseudogap structure in the DOS.  Similar results have been obtained earlier 
in Ref.~\cite{[8]}, where a $T$-matrix approximation was employed to treat 
disorder-induced resonant scattering. The above describes the competing 
tendency of the correlation-induced low temperature coherence scale, 
$T_{coh}$ in the $d=\infty$ Hubbard model, with the disorder-induced 
incoherence, which tends to suppress it, driving it eventually to zero, 
leading to a non-FL metal state.  A large enough disorder strength leads to 
band-splitting, as in usual CPA, leading to a continuous transition to a 
disorder driven insulating state.

\begin{figure}[htb]
\epsfxsize=3.4in
\epsffile{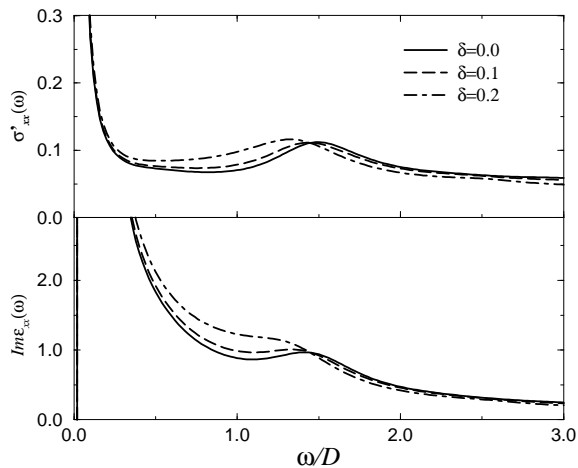}
\caption{Real part of the optical conductivity 
$\sigma_{xx}$ for $U/D=3.0$, $\delta=0.1$ (continuous) and $\delta=0.2$ 
(dot-dashed).  The lower panel shows the imaginary dielectric constant for
the same parameter values.  Note the isosbectic point in both quantities.}   
\label{fig1}
\end{figure}

Armed with this information, we are ready to discuss our results. We choose 
a gaussian unperturbed DOS, and $U/D=3.0$ to access the strongly correlated 
FL metallic state off $n=1$, ignoring the possible instability to an 
AF-ordered phase.  All calculations are performed at a low temperature,
$T=0.01D$.   We are mainly interested in the variation of the EELS lineshape 
with hole doping, given here by $\delta=(1-n)$.  This fixes the chemical 
potential, and the FL resonance position, and the IPT describes the evolution 
of spectral features in good agreement with exact diagonalization 
studies~\cite{[5]}.  In view of the ability of the IPT to reproduce all the 
qualitative aspects observed in $\sigma_{xx}(\omega)$, we believe that is a 
good tool in the present case. Fig.(1) shows the optical conductivity and 
the longitudinal dielectric constant.  $\sigma_{xx}(\omega)$ agrees with 
calculations performed earlier in all the main respects; in particular, it 
clearly exhibits the low-energy quasicoherent Drude form, the transfer of 
optical spectral weight from the high-energy, upper-Hubbard band states to 
the low energy, band-like states with increasing hole doping, and the 
isosbectic point at which {\it all} the  $\sigma_{xx}(\omega)$ curves as a 
fn. of filling cross at one point, to within numerical accuracy. It is 
interesting to point out that such features have also been observed in 
experimental studies~\cite{[10]}.  Correspondingly, Im$\epsilon_{xx}(\omega)$ 
also shows the isosbectic point, the explanation for which is identical to 
the heuristic one proposed recently by us for the case of 
$\sigma_{xx}(\omega)$~\cite{[11]}.

\begin{figure}[htb]
\epsfxsize=3.4in
\epsffile{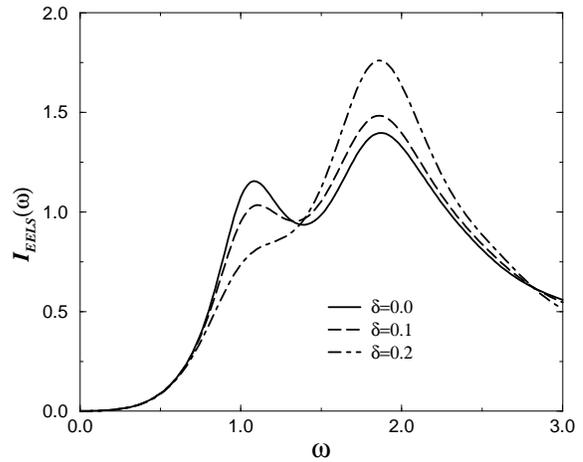}
\caption{  The electron-energy loss spectra for the doped Hubbard model for
$U/D=3.0$, and $\delta=0.1, 0.2, 0.3$ as shown.  The curve with circles denotes
the Drude contribution from renormalized single-hole excitations in the 
metallic phase.}
\label{fig2}
\end{figure}

\begin{figure}[htb]
\epsfxsize=3.4in
\epsffile{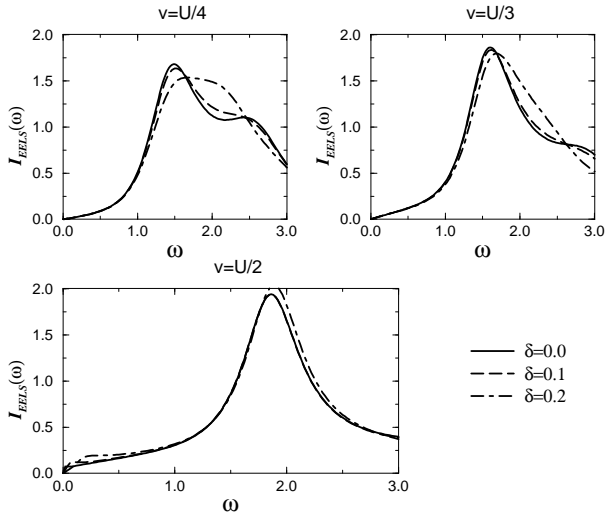}
\caption{  The EELS spectra for the $d=\infty$ Hubbard model for $U/D=3.0$ and 
$v/U=1/4, 1/3, 1/2$.  See the text for the explanation of various 
structures in $I_{EELS}(\omega)$ in the light of the underlying structure 
of the local spectral density of the HM in $d=\infty$.}
\label{fig3}
\end{figure}

In Figs.~\ref{fig2} and~\ref{fig3},  
we show the theoretically calculated EELS spectrum for $U/D=3.0$ 
and for different band-fillings as a function of disorder strength, $v$.
This allows us to access the intermediate correlation regime, where the
absence of any small parameter in the problem requires use of controlled 
methods that interpolate correctly between the weak- and strong coupling 
limits (such as the $d=\infty$ method used here).  From the calculation of the
optical response, we have checked that Im$\sigma_{xx}(\omega=0)=0$, implying
that ${-Re1/\epsilon(0)}=0$, as required~\cite{[12]}. With $U/D=3.0$, we 
observe a peak at $\omega \simeq U/2$, associated with the transitions from 
the lower-Hubbard band to the quasiparticle resonance. We associate this peak 
to the strongly renormalized particle-hole excitations in the strongly 
correlated metal.  In this context, two features are worthy of mention: at 
low energy, the loss intensity goes quadratically with $\omega$, and the 
particle-hole peak position shifts with increasing doping. 
The shift of this feature correlates with that of the mid-IR peak 
in the optical conductivity (Fig.(1)), and is thus a clear manifestation of 
the transfer of optical weight to lower energy with hole-doping.  Moreover, 
appreciable loss intensity is observed at higher energies ($\omega > 2.0$) 
as a smooth continuum, in contrast to the expectations from a weakly 
correlated (very small $U/D$) FL.  
 This clearly shows how the delicate balance between 
coherence and incoherence is controlled by the transfer of optical spectral
weight in the full EELS spectrum.  In fact, 
we expect the coherent feature in the full spectrum to become sharper as $T$ 
($0.01D$ in our analysis) is lowered further.  We see from the above that 
coupling to high-energy, incoherent, multiparticle excitations strongly 
renormalizes, but does not destroy, the coherent response in a strongly 
correlated Fermi liquid. However, the analysis presented above shows that 
a proper treatment of coherence (related to itinerance) and incoherence 
(coming from the local, atomic-like features) on an equal footing is necessary 
to obtain consistent results.  The EELS at very low energy are well described 
by a Drude fit, but features like the damped plasmon peak at 
$\omega/D \simeq 1.0$ are distinctly non-Drude-like, and can only be reliably 
accessed by the full analysis, as described above.

Given the above, we expect small coherence destroying perturbations to tilt 
the balance in favor of low-energy incoherence. Consideration of doping-induced
static disorder has precisely such an effect on the low energy EELS spectrum.  
Consistent with the disorder-induced smearing of the Drude part in 
$\sigma_{xx}(\omega)$, the coherent part in the EELS spectrum is replaced by 
a broad peak, which still shows remnants of the coherent peak at $v=0$ 
($v=U/4$).  However, the low-energy $\omega^{2}$ dependence in the pure 
case is replaced by a {\it linear} $\omega$ dependence in this case.
The complete destruction of the quasiparticle behavior in the DOS and the 
Drude-like response at $v=U/3$ is consistent with the complete destruction 
of the "coherent" peak in the EELS spectrum; we understand this as overdamping 
of the collective p-h peak by disorder-induced strong scattering as $v$ is 
increased.  In the strong disorder regime, the metallic state is incoherent 
with a pseudogap feature in the DOS~\cite{[11]}, and the action of the 
potential $V_{{\bf k}{\bf k'}}$ between the Hubbard band states does not 
create well-defined elementary excitations, leading to an incoherent response 
at low energy.

In experiments carried out on the cuprates in their normal state, a very 
broad plasmon peak is revealed~\cite{[12]}, implying poorly defined plasmon 
excitations in the correlated metallic state. These studies also reveal that 
the optic ``plasmon'' disperses quadratically (in $q$). Recent studies 
indicate an acoustic-like heavily damped plasmon mode at small 
${\bf q}$~\cite{[13]} with a {\it linear} $\omega$ dependence in 
$Bi_2Sr_2CaCu_2O_8$, alongwith a single broad peak.
We cannot directly compare our results to those of ~\cite{[13]}, the EELS 
experiment was carried out in the reflection mode, and this measures the
quantity $Im g(0,\omega)$, which is not directly related to what we have
calculated here.  However, looking at the results of ~\cite{[12]}, we see
that the essential features seen there are indeed reproduced by our 
calculation. To make contact 
between our calculations and experiments, we shall suppose that the one-band
Hubbard model is understood to be the {\it effective} model describing the
coupled spin-carrier dynamics in the Cu-O planes.  We will also assume that
bandstructure effects are not crucial to the case under study, and that the 
source of the anomalous behaviors is rooted in the dynamical effects of the
strong, local Hubbard interaction, an assumption well-supported at doping 
levels near the optimal~\cite{[14]}.  We stress that the above conditions 
will be modified in the pseudogap phase, where the precursor effects of 
$d$-wave pairing fluctuations-intrinsically non-local, are 
appreciable~\cite{[15]}.  In this doping range, the dynamical effects of 
non-local correlations will probably lead to even stronger deviations from 
a FL picture, though a concrete calculation including such effects remains to
be developed.  With these caveats in mind, we find good 
qualitative agreement between our results for $v=U/3$ of Fig.~\ref{fig3} and 
the experimental 
EELS lineshapes~\cite{[12]}; the quadratic (in $\omega$) behavior
at low $\omega$, the broad ``plasmon'' peak (strongly damped), the asymmetric
lineshape and the continuum response at higher energy are all reproduced in 
qualitative agreement with experiment.  It would be interesting to check 
whether the plasmon peak shifts to lower energies with hole doping; this 
would be an interesting check on the validity of the approach presented here.  
The comparison is quite good upto $\omega/D \simeq 3.0$, beyond which the 
simple single-band modelling does not apply anyway.   

There is extensive experimental work~\cite{[16]} indicating that the metallic 
phase above $T_{c}$ near optimal doping in cuprates is not describable in 
terms of Landau Fermi liquid ideas.  Prelovsek {\it et al.}~\cite{[17]} have 
recently considered the question of the EELS lineshape in the $t-J$ model, 
making use of results obtained from finite-temperature Lanczos techniques.
Within our approach, it is known that the metallic state off $n=1$ above the 
lattice coherence scale, $T_{coh}$, is not a Fermi liquid~\cite{[5]}; the 
calculation carried out above is then consistent with the non-FL charge 
dynamics if the $T_{coh}$ can be driven sufficiently low (below $0.01D$ used 
here).  We expect precisely a very low $T_{coh}$ near $n=1$ for $U/D$ near 
the Mott transition~\cite{[5]}. Comparing our results with those found 
in~\cite{[17]}, we observe that our results are in nice agreement with 
theirs, but the results presented here are more transparent, being based on 
an analytical scheme.  In addition, the results show that the $d=\infty$ 
approach is capable of producing the low- and high-energy features on the same 
footing, and moreover, permits us to include effects of doping-induced static 
disorder in a consistent way.  In view of this, we believe that the approach 
described here should also be applicable to a wide variety of doped transition 
metal oxides, where the combined effects of correlations and disorder are 
especially pronounced~\cite{[2]}.  Additionally, effects of orbital degeneracy,
etc, in real materials can be treated by a suitable generalization of the 
above procedure~\cite{[5]}.  This is a non-trivial problem, because the
solution of the impurity model is harder.  We hope 
to address this issue in future work. 

In conclusion, we have considered the energy loss function of a model 
for transition metal oxides, and captured the effects of strong, local
correlations and static disorder in a consistent way.  Comparison of our 
results with the experimental EELS spectra for cuprates in the ``normal'' 
state above $T_{c}$ near optimal doping shows that all the main observed 
features are consistently reproduced by our calculation.  

\acknowledgements
LC was supported by the Funda\c c\~ao de Amparo \`a Pesquisa do Estado 
de S\~ao Paulo (FAPESP). MSL thanks Prof. E. M\"uller-Hartmann for 
encouragement and advice and Prof. G. Sawatzky for a discussion of the 
experimental results. MSL was supported by the SfB 341 of the DPG.

\end{document}